Defining, Evaluating, Preparing for and Responding to a Cyber Pearl Harbor

Jeremy Straub
Institute for Cybersecurity Education and Research
North Dakota State University

1320 Albrecht Blvd., Room 258
Fargo, ND 58102

Em: jeremy.straub@ndsu.edu
Ph: +1-701-231-8196
Fx: +1-701-231-8255

**Abstract**
Despite not having a clear meaning, public perception and awareness makes the term cyber Pearl Harbor an important part of the public discourse. This paper considers what the term has meant and proposes its decomposition based on three different aspects of the historical Pearl Harbor attack, allowing the lessons from Pearl Harbor to be applied to threats and subjects that may not align with all aspects of the 1941 attack. Using these three definitions, prior attacks and current threats are assessed and preparation for and response to cyber Pearl Harbor events is discussed.

## 1. Introduction

Pearl Harbor is a seminal moment in U.S. military history and in U.S. history, more generally. Roosevelt proclaimed the day of the attack "a date which will live in infamy" [1] quite accurately, as nearly 80 years later this attack is still part of the national discourse and shaping U.S. national perception of present-day events.

The information technology revolution, which has resulted in a significant reliance on computers and computing equipment (such as Internet of things devices) in nearly every aspect of society, has raised the concern that new types of attacks are now possible. Whether carried out by criminals, terrorists or nation-states, these attacks may represent attempts to impair individuals' daily lives potentially as an attempt to directly achieve a political goal, to impair military readiness or as part of a combined cyber and conventional attack. Because of the surprise factor, the potential that a weaker force might use such an attack to target a stronger one and other potential similarities, these attacks have been commonly referred to as cyber Pearl Harbor attacks. Famously, former Defense Secretary Leon Panetta [2] cautioned that a future cyber-attack "could be a cyber Pearl Harbor; an attack that would cause physical destruction and the loss of life. In fact, it would paralyze and shock the nation and create a new, profound sense of vulnerability."

This paper considers the concept of a cyber Pearl Harbor. It reviews the 1941 attack and what was learned from it, in hindsight. It considers what characteristics a cyber version of such an attack might have, discussing the multiple prior uses and definitions of the cyber Pearl Harbor term. To facilitate analysis, the cyber Pearl Harbor concept is divided into three components. These components are used to consider whether attacks significant and relevant enough to evoke comparison to Pearl Harbor have already occurred and to define the characteristics that a future cyber Pearl Harbor attack may have. How to prevent, prepare for and respond to such an attack are all considered. Additionally, the application of the term to attacks against other nations is briefly discussed.

## 2. Background

A full understanding of the cyber Pearl Harbor term requires consideration of three elements, which are discussed in this section. First, the challenges posed by cyber threats and opportunities posed by cyber capabilities are discussed. Next, the historic Pearl Harbor attack is reviewed. Finally, the history of the cyber Pearl Harbor term is reviewed.

### 2.1. The cyber threat challenge and opportunity

The context for any discussion of cyber threats must be the benefits that computing technologies bring and the harm that their interruption, error, failure or unavailability can cause. Computing technology has dramatically changed the way that those in high-technology-use countries live and work. It has enhanced education [3] and been used to facilitate relationships between the elderly [4]. It has enhanced how health care is provided and its quality [5], reducing costs [6] and facilitating in-home treatment [7], in some cases. It has increased access for those with disabilities [8], improved tourism [9] and generally raised standards of living and improved individuals day-to-day lives [10].

Many of these benefits have come through inserting information technology into virtually every process in society. In some cases, information technology is augmentative with its removal or malfunction simply reducing capabilities. In other cases, the technology is integral to functionality so that devices, vehicles and even entire power grids can be rendered non-functional by its failure and individuals injured by its maloperation. Technology systems can be attacked by insiders [11], criminals [12] and even nation-states [13]. These attacks can stop or cause maloperation of medical devices [14] and otherwise interfere with patient treatment and impact patient safety [15]. They interfere with individuals' privacy [16] as well as with the devices that individuals and businesses use and the networks that transfer data [17]. Most problematically, the number of breaches and records breached is growing with 56% year-over-year growth in the first quarter of 2019 [18] and 4.1 billion records compromised through 3,800 reported breaches in the first half of 2019 alone [19]. While these numbers are staggering, they refer mostly to conventional breaches. Individuals have gotten so used to these breaches that they now pay limited attention to the notifications of them [20] and display symptoms of fatigue in managing their online privacy [21].

## 2.2. The Pearl Harbor attack

There is some dispute as to what Pearl Harbor actually was, in the context of World War II and American consciousness. It has clearly been memorable, as its continued use as a point of reference to the modern day [22,23] demonstrates. Panetta, when evoking Pearl Harbor in the cyber context, sought to convey the idea of a "a devastating digital attack on critical national infrastructure" and to "shake up ordinary citizens, to awaken them to the seriousness of the situation" [22]. He described the threat as one of having the "potential for a paralyzing attack" [22].

Some, though, would disagree with this characterization of Pearl Harbor. There seems to be little dispute that the attack was a surprise. Despite declining relations with Japan [24,25], the monitoring of Japanese communications [26] and breaking many of their codes [26], the attack was not predicted by U.S. intelligence. Even the signs that were present were dismissed due to a belief that the superiority of U.S. forces would deter this type of an attack [25]. Wirtz [25] notes that commanders at Pearl Harbor failed to heed "'war warnings', last minute indications of an approaching enemy, and even reports of an engagement with enemy forces" for this reason. Mueller [27] contends that, despite the surprise factor, the impact of the attack was a "military inconvenience" rather than a disaster as most of the damaged equipment was either already nearly obsolete or quickly repaired. Further, he suggests that the equipment production response quickly eclipsed what was lost to the attack [27]. The attack awoke the United States to the Japanese threat; however, this was (Mueller [27] contends) to the definite detriment of Japan and the potential detriment of the United States (which he suggests might have been better served by adopting a containment doctrine rather than the direct engagement that ensued).

Importantly, while the common perception of Pearl Harbor may be one of devastation, a more nuanced look may suggest that the event is best defined by the reaction that it produced. It unified and engaged the country in opposing Japan [27]. Mueller, though, suggests that this is the opposite of what Japan sought. He contends that the Japanese goal was to impair or eliminate the United States public's will to fight.

Thus, a Pearl Harbor-type event can be considered to be an event that is devastating and unexpected. It can be thought of as an attack that spurs anger and a desire to fight the enemy. Or, it can be thought of as an attempt to attack the enemy's will to fight. All three of these perspectives are key to consider when attempting to define a cyber Pearl Harbor.

## 2.3. A cyber Pearl Harbor

There has been significant discussion around the use of the cyber Pearl Harbor term. Discussion has ranged from who coined the term (with Hamre, Marsh, Bidzos and Schwartau all having some claim to its development and evolution [28] and Hamre, Deutch [28] and Panetta [23] involved in publicizing it) to the impact of its use on U.S. domestic and foreign policy.

Over time sentiment has changed significantly. In 1998, Smith [29] contended that an "electronic Pearl Harbor" was a form of hype and that many cyber threats were "the modern equivalent of ghost stories." A number of the actions he predicted as unlikely or unimpactful have occurred; however, even at present many regard the impact of cyber-attacks as minimal, with Lawson and Middleton [28] recently arguing that, referring to Russian cyber-attacks as part of their Crimea annexation, "the immediate effects of these attacks were limited, as have been the effects of cyber-attacks on the shape of the military conflict more generally." Since 1998, there have been thousands of data breaches compromising millions and millions of records of data. The Equifax breach [30] alone revealed information for over 100 million Americans. Clearly, cyber threats cannot be dismissed any more; however, their prospective use and precise efficacy in warfare still remains a subject of significant debate.

While Lawson and Middleton [28] contend that the cyber Pearl Harbor term "has had a largely stable meaning" which they say is of "catastrophic physical impacts from cyber-attacks on critical infrastructure" other uses are prevalent. Wirtz [25] provides an aligned, but more nuanced definition where a "militarily inferior opponent will utilize a cyber-attack in conjunction with conventional military operations to present the United States with a fait accompli" which might be a surprise both from a timing and technical perspective. In any case he notes that it will "incapacitate extraordinarily powerful forces developed and deployed at enormous expense" with an attack "undertaken at relatively minimal cost" [25].

Not all uses conform with Lawson and Middleton's definition, however. Smith's [29] 1998 "electronic Pearl Harbor" article describes a scenario where "blackouts occur nationwide, the digital information that constitutes the national treasury is looted electronically, telephones stop ringing, and emergency services become unresponsive" without invoking "physical impacts."

Loui and Loui [31] suggest, instead, that the term is "a reminder of the risks of feeling invulnerable and being unprepared for – even complacent toward or doubtful of – an attack of this scale and nature." They also discuss [31] how the term could relate to a "vulnerability of a system's defense, particularly when faced with a paradigm shift." Smith and Erickson [32], alternately, suggest that "the term 'Pearl Harbor' has come to denote the concepts of an infrastructure left completely undefended and how only a massive attack makes society take that exposure seriously."

Lawson and Middleton [28] explain that the definition and use of the term is important due to "the importance of language, including metaphor and analogy, for correctly framing and responding to national security threats." They contend that the term has had "real, negative impacts" as it is "not merely used by officials in public speeches" but also in "internal cyber security discourse and strategizing" and has "distracted us from the real threats we face" [28]. They suggest that the term may best be discontinued or perhaps redefined to better align national interest with real threats.

## 3. What would define a cyber Pearl Harbor?

There have been numerous suggestions as to what a cyber Pearl Harbor is. These range from the large catastrophic attack discussed by Smith [29] to Wirtz's [25] combined cyber and conventional warfare attack to Lawson and Middleton's [28] suggestion of it best being used to describe "many small attacks, or even large attacks, that occur but somehow go unnoticed." Clarke [28,33] has even suggested that the metaphor might be evoked as "daily digital Pearl Harbors."

The problem with all of these approaches is that they fail to separate the elements of the Pearl Harbor attack: attacker intent, the actual attack event and the impact of the attack. While the historic Pearl Harbor attack, of course, had specific characteristics for all three, a modern cyber-attack might have characteristics similar to only one or two of the three. Thus, a comparison to the Pearl Harbor attack might be warranted (in that it is useful for aiding understanding or conveying impact); however, more nuance in the use of the term could aid in identifying what parts of the metaphor were being invoked.

For the purposes of the analysis herein, and more generally, the cyber Pearl Harbor analogy is decomposed into three parts: goal, event and impact. This decomposition is depicted in Figure 1 and each component is now discussed.

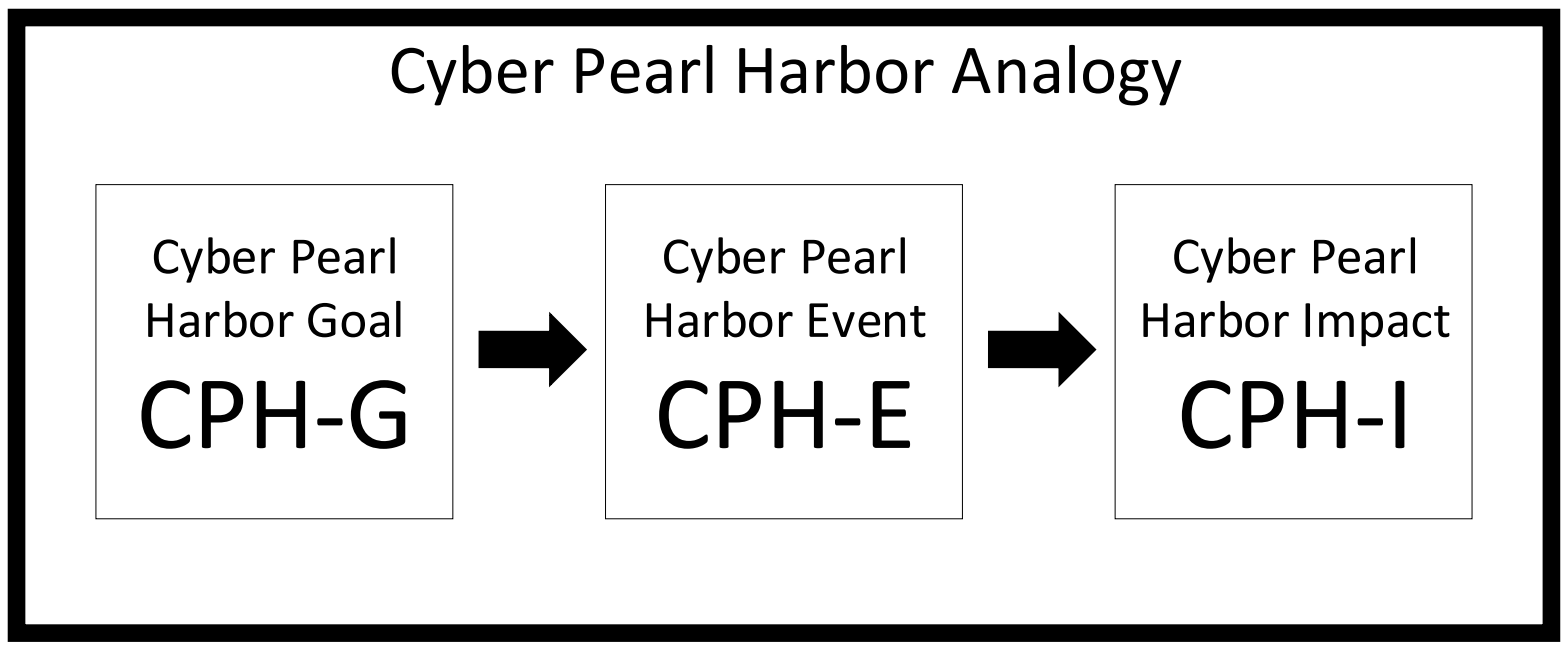

Figure 1. Cyber Pearl Harbor analogy components.

The cyber Pearl Harbor goal (CPH-G) term refers to the intent of the attackers. Historically, Mueller [27] notes that this was "inflicting psychological shock," turning public sentiment against the war and destroying the U.S. will to fight. Thus, CPH-G can be defined as an adversary having a goal of deterring future conflict due to the fear and/or damage caused by a strike.

The cyber Pearl Harbor event (CPH-E) term refers to the actual event itself. Historically, this was the attack on the Pearl Harbor base. There are multiple lessons that can be derived from this historical attack: the importance of intelligence, not disregarding information because of a believed superiority and even how actions taken in response to one perceived threat can make another more devastating when it actualizes. The CPH-E term will be defined as an event that though perhaps projected, still has a surprising and large-scale outcome.

Finally, the cyber Pearl Harbor impact (CPH-I) term refers to the actual result of the attack. Historically, this was mobilizing and unifying America in support of the war effort. As Mueller [27] notes "the attack on Pearl Harbor was phenomenally successful in its shock effect, but the shock was exactly the opposite of the one the Japanese hoped for." Thus, the CPH-I term will refer to the impact of an event that awakens the public (or groups) to the need to take action in response to cyber threats.

4. Has a cyber Pearl Harbor or near-cyber Pearl Harbor occurred?

In this section, seven incidents will be reviewed to ascertain whether they meet the standards as a CPH-G, CPH-E or CPH-I. These seven incidents are the Yahoo data breach, the Office of Personnel Management (OPM) breach, the Equifax breach, the Sony breach and the Marriott breach as well as the Stuxnet attack and the Russian invasion of Crimea. Six of the seven are on CSO Magazine's list of the 18 largest data breaches [34]. First, each of the seven incidents is described. Then, in the subsections that follow, each of the CPH types is more thoroughly defined and each of the seven incidents is assessed relative to the presented definition.

The Yahoo breach [35] is the most impactful breach identified by CSO Magazine [34] and is actually a combination of breaches that exposed 3 billion records [36]. The series of attacks is believed to have started in 2013. The theft of 500 million records was discovered in 2014. An earlier attack was discovered in 2017, where it was initially believed that approximately one-billion records were stolen in 2013 [35]. Yahoo admitted a failure to properly investigate the 2014 breach [35], which it believed was targeted at 26 users and was conducted by a "state-sponsored hacker" [35]. Data stolen included password hashes, email addresses and phone numbers [35].

The OPM breach is likely one of the best documented data breaches in history, as the subject of a congressional report [37] of over 200 pages. While the breach was far smaller than many others, affecting only 22 million individuals' records, the stolen information included 21.5 million background investigation data files [38], which are perhaps the most robust caches of personal information in the United States. There is also derivative impact, as these files contain names, addresses and other information for relatives and other associates of the individuals whose data was compromised. This attack was believed to be perpetuated by the Chinese Government [38] and has the potential use of identifying those with access to classified U.S. material, their associates and their personal secrets. It has also been suggested [37] that these details could be used to recruit U.S. intelligence operatives for the Chinese Government and to identify possible U.S. operatives in China through association and co-location. The breach started in 2012 and was not detected until 2014 [38]. Another related breach occurred in 2014 which was detected in 2015 [38].

The Equifax breach resulted in the theft of information for 148 million individuals [39]. The breach occurred between May and July of 2017 and resulted in the disclosure of social security numbers, birth dates, addresses and some credit card numbers, driver's license numbers and dispute documentation [39]. The exact details of the attack remain unclear; however, Terrasi [40] claims that it is "the most serious in history" due to a "grossly inadequate response" including a delay in the disclosure of the issue. Martindale [41] notes that the attack was responsible for "severely compromising the identity of hundreds of thousands" and "left the majority of the country vulnerable to fraud."

The Sony breach is included as it was called a cyber Pearl Harbor by the National Security Agency director [28]. However, it didn't make Terrasi's top five list [40] or Armerding's top 18 list [34]. The Sony attack included a number of unusual elements. First, the company appears to have been attacked by a nation-state (North Korea) to stop the distribution of a film [42] (i.e., a form of speech). Second, the United States Government directly intervened, threatening action and potentially cutting North Korea off from the Internet for a period of time [42]. Third, Sony directly and aggressively responded, conducting denial of service attacks on those with its content and posting files online to attempt to misdirect those searching for the stolen content [42]. Wolff [42] contends that the Sony attack was notable due to the publicity it generated at the time, the U.S. government involvement on Sony's behalf and Sony's own actions in response.

The Marriott breach has been argued [43] to be an action of the Chinese Ministry of State Security as part of an attempt to gain information about “executives and American government officials with security clearances.” In this regard, Sanger, et al. [43] state that it is related to attacks on health insurance companies and the OPM breach. The stolen data includes identifying information, credit card numbers and passport numbers. Monaco proffered that the passport details “would be particularly valuable in tracking who is crossing borders and what they look like” while Lewis argued that they were “seeking to identify American spies – and the Chinese people talking to them” [43]. Alperovitch goes even further suggesting that the data “can be used for counterintelligence, recruiting new assets, anticorruption campaigns or future targeting of individuals or organizations” [43]. Armerding [34] identified the breach as the second most impactful of all time.

Though Armerding [34] considers it to be only the fifteenth most impactful breach of all time, Stuxnet was an attack by two nation-states on a third. It was also significant because it damaged or destroyed not only physical equipment but equipment that was being used to create munitions. Armerding [34], further, notes that it will “serve as a template for real-world intrusion and service disruption of power grids, water supplies or public transportation systems” in the future. Stuxnet was developed to prevent or delay Iranian nuclear weapon capability development [44]. To do this, it targeted particular Siemens programmable logic controllers in a particular configuration and has minimal impact on computers without these controllers. The Stuxnet worm was developed by U.S. and Israeli intelligence agencies to target Iranian centrifuges [44]. Not only did it alter commands to the controllers, it also masked itself by sending false data about controller / centrifuge performance, making everything look like it was working until the centrifuge was damaged. The program was code-named Operation Olympic Games and was designed to prevent the need for Israeli airstrikes against Iranian nuclear facilities (if it appeared that Iran was close to developing a nuclear weapon) which could start a war in the region [44].

The Russian invasion of Crimea and other Russian activities in the Ukraine are perhaps the most pronounced examples of the use of cyber capabilities in support of real-world attacks. In the Crimea invasion, Russian forces utilized a combination of conventional forces, cyber-attacks and social engineering [45] to facilitate their annexation of the region, which already had strong Russian support. This move was part of a broader Russian strategy to regain prominence within the global political landscape and secure its leading position within the former USSR region [46]. This campaign has included two cyber-attack induced blackouts [46] as well as a number of more conventional attacks against websites and software systems [45]. These attacks follow a pattern of cyber-attacks in support of Russian positions that has included previous similar efforts in Estonia and Georgia [46]. In the Georgia attack, cyber and conventional attacks were coordinated, and the cyber-attack supported the objectives of the conventional one [46]. According to Greenberg [47], cyber activities since the Crimea annexation have “systematically undermined practically every sector of Ukraine: media, finance, transportation, military, politics, energy.” Geers noted that “you can’t really find a space in Ukraine where there hasn’t been an attack” [47]. Ukraine has become such a “test lab” that cybersecurity professionals for around the world – both from the private sector and governments – are setting up presences there to analyze attacks before they see them closer to home.

It is important to note that these are contemporary incidents and that additional – and potentially corrected – details may emerge over time. Given this, the assessment that follows is based on the current understanding of the incident based upon the cited sources. If details of the incident change, this analysis still serves to demonstrate the CPH-G, CHP-E and CPH-I classifications, though the appropriate classification for a given incident may change. Table 1 presents and overview of the classifications and the following subsections describe the classifications in more detail.

Table 1. Overview of Classification.

| | CPH-G | CHP-E | CPH-I |
|---|---|---|---|
| Yahoo data breach | | M | |
| Office of Personnel Management | M | Y | |
| Equifax breach | | | |
| Sony breach | Y | | |
| Marriott breach | | | |
| Stuxnet attack | Y | Y | |
| Russian invasion of Crimea | Y | Y | Y |

Key: Y=Yes, M=Marginal

4.1. Cyber Pearl Harbor Goals

CPH goals mirror the goals of the Pearl Harbor attack. Namely, CPH-Gs are to create a deterrent against to the attack recipient's future actions. With the Pearl Harbor attack (as described in Section 2), the Japanese intended to deter United States involvement both through the destruction of warfighting capability and through creating fear about the cost of engagement in the conflict. Note that goals do not need to be realized in order to be classified as CPH-Gs, they simply need to be stated or implicit objectives of the attacker.

The Yahoo data breach, Equifax breach and Marriott breach do not have currently known deterrent goals. Published reports (including those described earlier in this section) do not identify, for example, that attackers sought to deter users from using Yahoo or prospective guests from staying at a Marriott hotel. Equifax's primary business is not providing services to consumers; however, there was also no suggestion that attackers sought to deter businesses from using their services in the future.

Conversely, the Sony attack was thought to be a direct attempt to deter the company from releasing (and retaliation for releasing) a movie critical of North Korean leaders. The Stuxnet attack was designed to prevent or delay Iran's development of nuclear weapons. The Russian invasion of Crimea is more nuanced; however, the cyber attacks had at least the impact of impairing Ukrainian warfighting efforts, if not also deterring action to retake the Crimea region. Finally, the OPM breach may have a deterrent goal of knowing that an adversary has access to all of the personal information taken. This information could prospectively be used in the future (or perhaps even already has been used) for blackmailing, coercing or recruiting compromised individuals as spies.

4.2. Cyber Pearl Harbor Events

CPH events mirror the Pearl Harbor event in that they are defined as having characteristics including being unexpected (perhaps despite their being indications that they could happen), paradigm shifting and damaging to operational capabilities. Examples of a CPH-E in the cyber space include the development of a new exploit (perhaps based on a known vulnerability) as well as attacks against new targets and new types of targets and attacks that utilize new techniques. Attacks may also be CPH-Es because of the scope or scale of the attack, when the scope/scale is unexpected.

The Stuxnet attack introduced a new paradigm of attacking the physical world and air-gapped facilities through computer viruses and can be classified as a CPH-E for several reasons. In particular it is a paradigm-changer in that it shows that equipment of a type and in a configuration that was previously thought to be safe is actually vulnerable.

The Crimea invasion is similarly a paradigm changer through showing the effectiveness of the integration of cyber and physical world attacks which support each other. It also has the similarity of having the attacker's goals and ambitions known in advance but seeming unlikely for action to be taken to achieve them.

The OPM attack can be marginally classified as a CPH-E due to the information that was taken and being a direct and consequential attack against the United States government. Given the potential for retaliation from the United States, there were likely many that did not expect an adversary to be this brazen. Due to the significant impact on national security, there was a real potential for retaliation (including a physical world retaliation) that would not be invoked by most other attacks

The Yahoo attack can be marginally classified as a CPH-E due to its sheer size and the magnitude of data compromised. Beyond this trait, the attack lacks other CPH-E characteristics. The Equifax, Sony and Marriott breaches, while all impactful, lack unexpectedness, a paradigm shifting nature and significant technical advancement and are not classified as CPH-Es.

### 4.3. Cyber Pearl Harbor Impacts

CPH impacts are born out of incidents that awaken the general public to a new threat or to a threat that they were previously not concerned about and/or galvanize the public in support of an action to respond to, retaliate against or resolve the issue. It is important to note that a CPH-I does not need to have an attacker intend to awaken the public or generate support. In fact, the CPH-I may be exactly the opposite of what an attacker sought (if it had CPH-Gs).

The Crimea annexation is a good example of this as, within the region, it demonstrated the Russian threat and triggered a significant response. While the cyber element is only a part of what caused the public to become aware of and support actions to respond to the attack, it is non-the-less an example of a CPH-I

While the Marriott, Equifax, Yahoo, OPM and Sony breaches all attracted some public attention, none drove public awareness or interest in responding in a way similar to the Pearl Harbor attack. The Stuxnet attack clearly generated awareness in security and computing circles; however, it only minimally entered the public consciousness and has had nowhere near the level of impact as the Pearl Harbor attacks.

### 4.4. Summary

By defining the cyber Pearl Harbor incident classification into three sub-classifications, more nuanced evaluation, preparation and response can result. An incident could be described as an across-the-board CPH or one of the CPH categories, which is most appropriate, could be identified as its classification (instead of just the general CPH classification). There is a simple definition for what is required to be classified as a CPH-G, CPH-E or CPH-I presented in Table 2.

Table 2. Cyber Pearl Harbor sub-classifications

| | |
|---|---|
| CPH-G | Goal of launching an attack with an intent to prevent, limit or delay the target entity or country in carrying out its mission and tasks |
| CPH-E | An event which is unexpected, damaging and potentially includes and uses a technical innovation. |
| CPH-I | An incident that has CPH-I characteristics will have the result of increasing public interest in cybersecurity and / or a the public's willingness to devote resources to retaliating, responding or taking action to prevent future similar attacks. |

## 5. Cyber Pearl Harbor scenarios for other nations

While Pearl Harbor doesn't have the same place in the national consciousness for other nations around the world, the core concept of CPH-Gs, CPH-Es and CPH-Is is equally applicable.  In fact, one of the best examples of an incident combining both CPH-Gs and a CPH-E is the Russian invasion of Crimea in the Ukraine.  While the cyber aspects of the Crimea invasion were in combination with and in support of a human invasion and occupation, the Crimea invasion also has elements of CPI-Is.  These are not to the same level, necessarily, as the actual Pearl Harbor impact and the cyber aspects of the incident contribute only a fraction of the impact.

Given this, though the terminology used to describe it may not be the same, CPH-Gs, CPH-Es and CPH-Is can occur anywhere.  Moreover, identifying goals, events and impact of this type could be a useful tool, in many cases, to understanding what has occurred and its implications and either understanding or predicting (if the analysis is conducted during the incident) reaction.  In light of this, CPH-G, CPH-E and CPH-I based analysis may also be useful for those planning a cyber-attack to help them shape what impact an attack might have.

## 6. Preparing form responding to and preventing a cyber Pearl Harbor

When discussing preparing for, responding to, and preventing a CPH, the considerations vary by whether one is considering a CPH-G, CPH-E or CPH-I.  A nation state would ideally prevent other nations from having CPH-Gs which could lead to CPH-Es.  States would also seek to prevent CPH-Es in their own right, in case a nation attacked due to CPH-Gs that were not prevented or for a non-CPH-G reason.  Conversely, nation states will typically look to prevent CPH-Is in their opponents – not their own – populations.  Each is now discussed and the relationship between the different CPH types and activities is depicted in Figure 2.

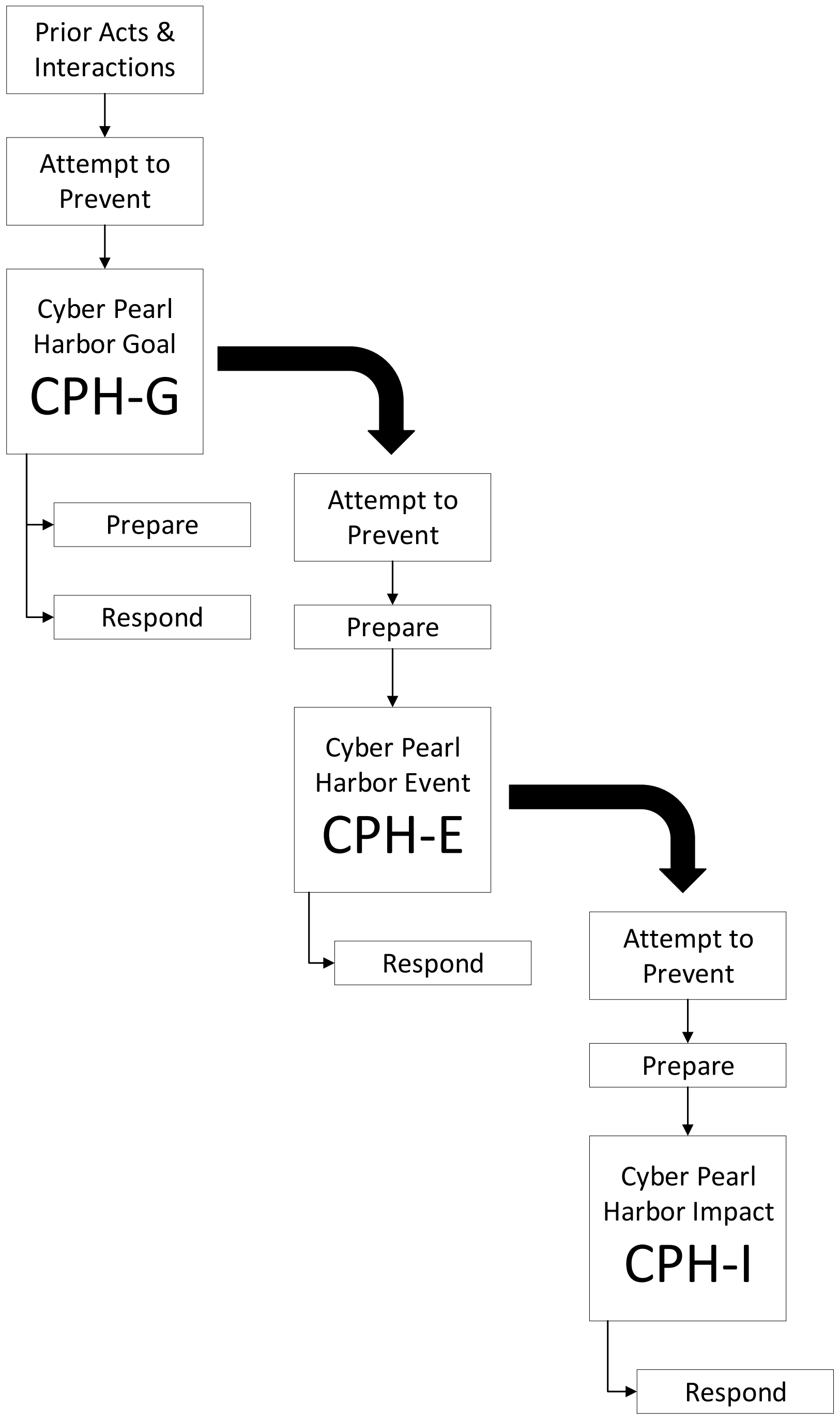

Figure 2. Relationship between CPH-G/-E/-I and preparations, prevention and response activities.

6.1. Preparing

Preparation for a CPH incident will vary dramatically based on whether one is preparing for a CPH-G, CPH-E or CPH-I. Relative to an adversary with CPH goals, preparation may be multi-faceted. First, this may consist of activities designed to identify adversaries that may have CPH-Gs towards the nation through intelligence and diplomatic activities. Second, preparation may include taking actions designed to reduce the desire of the prospective adversary or adversary to have these goals against the state (though this may fall more into the preventing category, discussed below). Third, preparations will need to include the activities required to prepare to conduct response activities if a CPH event or a non-CPH-E in support of CPH goals occurs. It is important to note, in this regard, that an adversary with CPH goals may express these in ways other than a CPH event. These may lead to a CPH-E or take other forms.

For an adversary likely to conduct a CPH event, preparation will focus on activities designed at prevention (discussed below) as well as to facilitate response to the CPH-E. Note, in this regard, that the very nature of CPH-Es makes preparation for response difficult. A CPH-E is characterized by being unexpected (but possibly with activities that may be detectable in advance as a warning). General preparedness for cyber-attacks, as opposed to preparing only for the things that are deemed most likely, may be the most effective strategy for CPH-E preparations. A state would also benefit from broadening its specifically targeted preparations to include preparations for events that seem less likely, but which would have significant strategic benefit to an adversary or prospective adversary. This is particularly necessary for powerful states where an adversary may see a surprise attack as the only potential advantage that it has over the larger state's more robust offensive and defensive capabilities [25].

The Pearl Harbor attack also demonstrated that preparations for one potential type of attack or incident can create vulnerabilities to others. In the historical attack, the movement of the fleet and orientation of aircraft made the United States more vulnerable to the attack than would have existed without these actions [25]. Similarly, developing network defenses and security mechanisms for one sort of planned cyber attack (or combined cyber and physical attack) may create opportunities for an adversary to exploit to launch an alternate type of an attack. Thus, preparations should be critically assessed for other vulnerabilities that they may create.

Preparations for CPH impact could take two forms. First, a state may wish to consider what the result of an attack causing CPH-I in its own citizens would be. Conceivably, this may be undesirable if it drags a nation into a drawn out, expensive and even potentially unwinnable war. Preparation, in this case, may be to prepare the public for an incident to attempt to mitigate a CPH-I from occurring or to channel the public response to it. Alternately, a state looking to go to war (or engage in conflict with an adversary) may prepare to channel the public anger of a CPH-I into support for this effort.

Second, a state may wish to prepare for CPH-I in its adversary. Part of these preparation activities may be attempting to prevent or reduce the likelihood of a CPH-I occurring (either in response to a planned or inadvertent attack against the adversary). This could potentially be accomplished through a propaganda campaign. Preparation for dealing with an adversary with an engaged population in support of a conflict against the state would be another potential scenario to prepare for. This preparation could include increased defenses as well as preparation for attacks from individuals and groups that are loosely affiliated with the adversary, in addition to the state adversary itself.

### 6.2. Preventing

Similar to preparing, preventing CPH-Gs, CPH-Es and CPH-Is varies significantly. The prevention of an adversary developing CPH-Gs can be accomplished in multiple ways. First, intelligence, propaganda and democracy efforts can be used to prevent the adversarial relationship necessary to foster CPH goals from developing. National policies that could potentially lead conflict could also be avoided. A variety of studies have suggested, for example, that U.S. policies have led to the development of current adversaries [48–50]. While accurately predicting the outcomes of every effort is difficult, if not impossible, efforts can be made to maintain relationships that prevent hostility from developing. Trade, for example, has been identified by many (e.g., [51]) as a way of reducing the likelihood of hostilities.

Alternately, a state can position itself to have response and retaliatory capabilities that deter the execution of activities in support of CPH goals, if not the formation of the goals themselves.

Prevention or deterrence of CPH events is similarly two-fold. First, a state can prepare its defenses and conduct intelligence activities to remove opportunities for successful adversary action and increase the likelihood of becoming aware of the activity in the planning stages. Second, a state can develop retaliatory capabilities that deter an attack due to fear of the state's response. This is the foundation of the mutual assured destruction paradigm [52] that has, so far, been successful in preventing significant conflict between nuclear powers. This same concept has been applied to cyber conflict [53,54].

Preventing or deterring CPH impacts, like preparation for them, must consider CPH-I in both adversaries and a state's own citizens. Preventing CPH-I in an adversary can combine attack development (such as to avoid targets and levels of impact that would have a galvanizing effect), propaganda and diplomatic efforts. CPH-I in an adversary can potentially also be prevented, or the formation of it deterred, by possessing a significant enough fighting and retaliatory capability such as to make an attack against the state seem unlikely or impossible to succeed. This can prospectively be accomplished via actual capability development and propaganda to project awareness of the strength of the state's capabilities. Notably, the Japanese had hoped that the U.S. would be deterred by the Pearl Harbor attack, not galvanized into action by it; however (as discussed in Section 2) this was a critical miscalculation.

Preventing or deterring CPH-Is in a state's own population may be desirable if the state's leaders do not desire to go to war (potentially because they do not see a way to win it, or for other reasons). This can be accomplished, prospectively, through the use of intra-state propaganda, before an attack, and defusing activities, after an attack.

### 6.3. Responding

Responding to CPH-Gs, CPH-Es and CPH-Is is also quite different. In the case of CPH-Gs, response is really a preparation or prevention activity. There is no specific response to an adversary having CPH goals, other than to attempt to dissuade them from holding them or prepare for an attack in support of them achieving the goal objectives.

For CPH events, response is a paramount consideration. This response will typically have multiple components. These, in most cases, will include immediate response activities that prevent additional damage, tend to any human injuries and physical damage and attempt to restore computer system functionality. This, in many cases, will need to be followed by a rebuilding phase where more permanent solutions to the damage are implemented (e.g., restoring the full functionality of affected systems and restoring them to their full pre-attack capabilities). Also, mitigation activities against future attacks should be undertaken based on the knowledge gained from analysis of the attack vector used, system vulnerabilities and the attack's results. These phases follow typical response patterns to natural disasters [55], as illustrated in Figure 3. In addition, retaliation activities will typically be warranted and desired in response to a CPH-E, particularly if it creates a CPH-I. The exact nature of the response and retaliation activities will obviously vary from event to event and circumstance to circumstance.

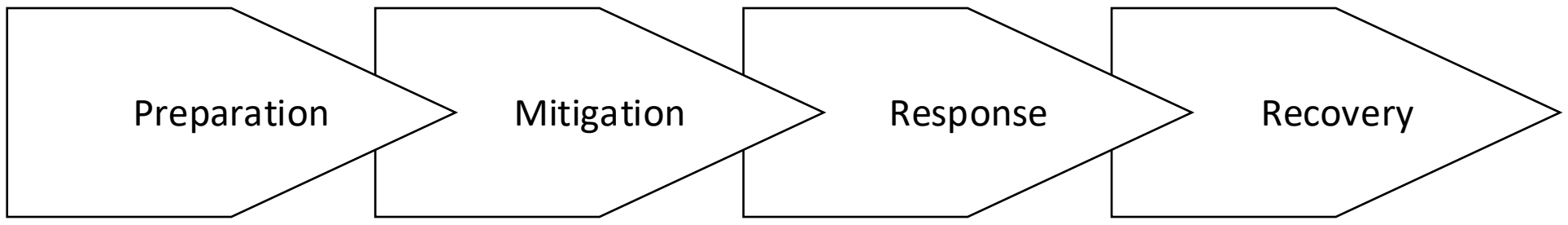

Figure 3. Model of a four-phase approach for dealing with emergency management.

Response for CPH impacts will typically vary significantly by the circumstance. If a state's own citizens experience a CPH-I then it is likely that this will lead to some sort of response (such as an attack on the adversary) action, similar to the responses for a CPH-E. Note that a CPH-E can occur without producing a CPH-I; however, it would be very difficult for a CPH-I to occur without a CPH-E. One possible scenario is where numerous smaller attacks over time combine to create CPH-I without a single large CPH-E incident.

If an adversary's citizens have CPH-I, then the response action required would be to prepare for and defend against the likely attack that will ensue. Again, the exact nature of such a response is highly situation dependent.

7. Using the CPH-G/-E/-I classification system

The utility of the CPH sub-classifications (goal, event and impact) is that they can be used as a model for preventing an adversary from producing CPH results or a model to create a CPH style impact against an adversary. In particular, a state that has CPH goals, may wish to launch an attack that is a CPH event without causing the CPH impact that could lead to extreme retaliation.

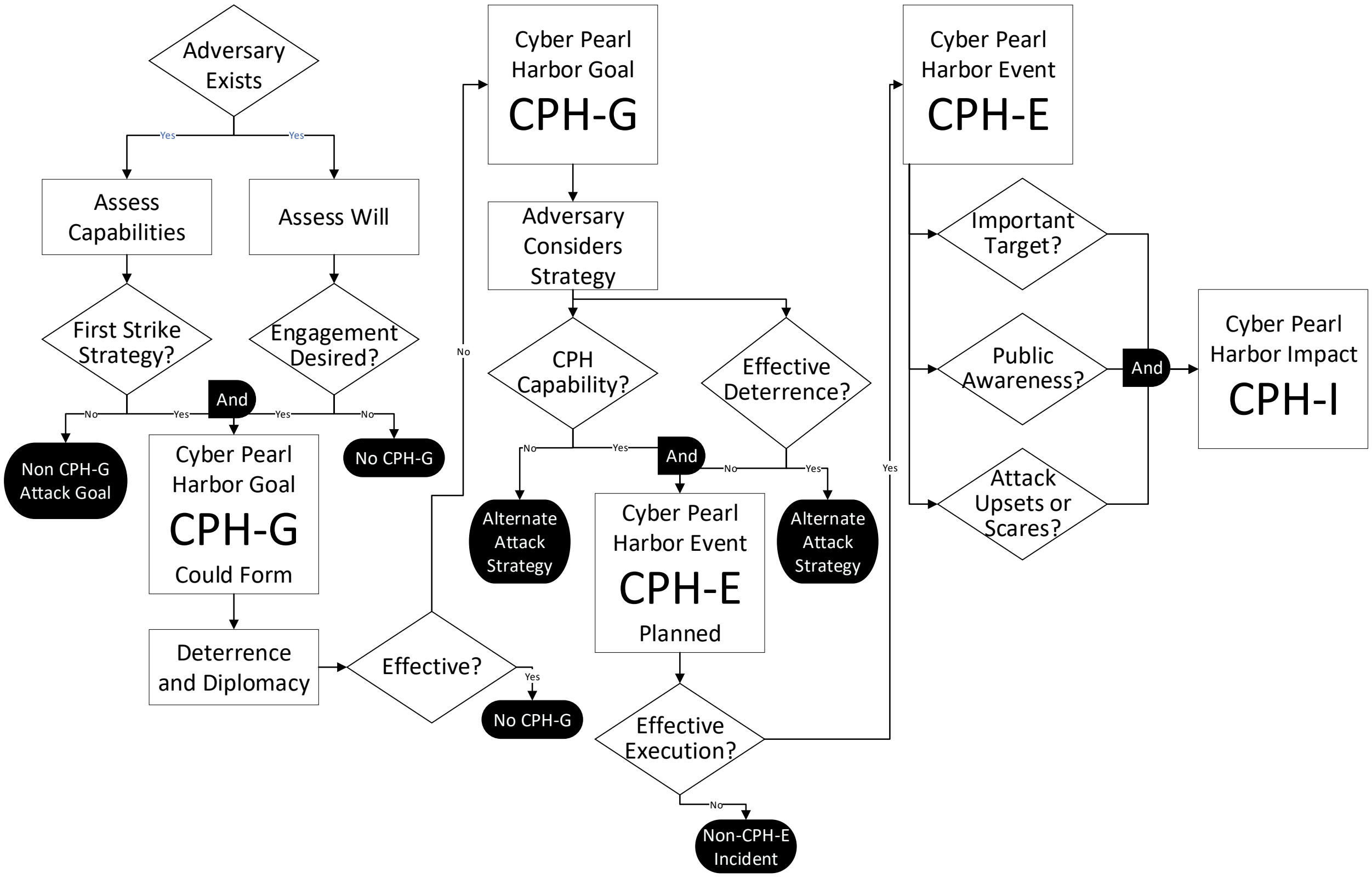

Figure 4. Pathways Including CPH-G, CPH-E and CPI-Is.

As another example, upon the occurrence of a CPH-E, a state may desire to create CPH impact, if galvanizing the public to support a war effort is desirable. Alternately, it may seek to prevent this from occurring if it does not plan to enter conflict or prefers a smaller-scale engagement. A state may also seek to launch an attack, which has CPH-E characteristics, against an enemy and, previously, simultaneously or subsequently, take action to mitigate the changes that it will create CPH impact.

Figure 4 depicts how an adversary, or a state, can move from a pre-engagement status to having CPH goals to deploying an attack that creates a CPH event to creating CPH impact. Notably, there are multiple decision points along the way (even in this simplified model) that can prospectively prevent CPH-Gs, CPH-Es and CPH-Is, providing an opportunity for both the state and its adversary to encourage the occurrence of these or their prevention as desired and dictated by their respective strategies.

## 8. CPH-G/-E/-I and considerations of attribution and non-state actors

Attribution and non-state actors impact the use of the CPH-G/-E/-I classifications and decision-making model. Attribution considerations can make deterrence problematic. If a would-be attacker believes they can achieve their goal without the attacked state being certain enough of who attacked to retaliate, virtually all deterrence-based prevention and response techniques are no longer applicable. Additionally, even temporary delays in attribution determination can impair the deterrence value of retaliation (as the attacker would potentially be able to achieve subsequent goals during this period). The potential for so-called 'false flag' attacks, where the attacker pretends to be another adversary (or an entity that was not previously thought to be an adversary) can be very problematic as they can remove the deterrence of retaliation and actually trigger retaliation by the incorrectly-retaliated-against party.

Attribution limitations can also make it difficult to conduct intelligence or to determine what entity may be planning a CPH-E or may have CPH-Gs, based on intercepted communications or other signals. This can, thus, impact preparation and prevention activities and prospectively reduce their effectiveness.

Non-state actors can also be very problematic as it can be difficult to determine the membership (which may be frequently changing) of loosely affiliated groups, which can impair intelligence activities and thus preparation and prevention. The lack of a target (i.e., because non-state actors don't have infrastructure or territory to attack) may also impair retaliation-based deterrence and response activities.

Both attribution and non-state actors present unique considerations. Their impact on CPH scenarios and the prospective role of non-state actors in state versus state conflicts remains a key topic for future study.

## 9. Conclusions and future work

This paper has discussed the concept of a cyber Pearl Harbor. The term, which springs from the Pearl Harbor attack by Japanese forces that led to U.S. involvement in World War II, denotes different things depending on its use, context and the individual using it. Herein, the term was presented as having three meanings, related to goals, events and impact. A definition for each was supplied and the three classifications were used to assess seven contemporary cyber attacks to identify if each had a goal, event type or impact similar to the Pearl Harbor attack. The relevance of the three classifications for planning and preparing for, responding to and attempting to prevent attacks was discussed. From this analysis, it was shown that the strategic and military lessons of Pearl Harbor are still relevant today and are very relevant to the cyber domain, if somewhat different in certain details due to the cyber context.